\documentstyle[twoside,fleqn,espcrc2,epsf,rotate]{article}

\newbox\rotbox

\newcommand{\AmS}{{\protect\the\textfont2
  A\kern-.1667em\lower.5ex\hbox{M}\kern-.125emS}}

\title{Negative-Parity Baryon Spectroscopy\thanks{Talk presented by
Frank X. Lee.}}

\author{Frank X. Lee\address{Department of Physics,
        The George Washington University, Washington, DC 20052
	and\\
	Jefferson Lab, 12000 Jefferson Avenue, Newport News, VA 23606, USA}
        and 
        Derek B. Leinweber\address{Department of Physics and
        Math. Physics, University of Adelaide, Adelaide, SA 5005,
        Australia}}
       
\begin{document}

\begin{abstract}
Results are reported for the first calculation of the low-lying
spin-1/2 odd-parity octet baryon masses using ${\cal O}(a^2)$ improved
gluon and fermion actions.  Methods for removing even-parity
ground-state contaminations from the two-point correlation functions
at zero and finite momenta are outlined.  We investigate the
properties of two odd-parity interpolating fields based on the
established interpolating fields for the nucleon ground state.
Isolation of the lowest-lying odd-parity state appears to be
sufficient to begin exploring odd-parity $N \rightarrow N^*$
electromagnetic transition form factors.
\end{abstract}

\maketitle

\section{INTRODUCTION}

  It is well recognized that the investigation of processes involving
the production of excited states of the nucleon ($N^*$) can provide
relevant information on the non-perturbative regime of QCD. In the
next years, the issue of $N^*$ physics will be addressed by planned
experiments at Jefferson Lab, which will provide new experimental data
in the resonance region with unprecedented accuracy.  There is
increasing demand for theoretical understanding of the $N^*$
properties from first principles calculations.  

In this work, we carry out exploratory calculations for odd-parity
spin-1/2 ($J^P=1/2^-$) baryon resonances on the lattice using ${\cal
O}(a^2)$ improved actions.  We utilize the ${\cal O}(a^2)$ tree-level
tadpole-improved Wilson gauge action and the ${\cal O}(a^2)$
tree-level tadpole-improved D$\chi$34 action originally proposed by
Hamber and Wu \cite{Hamber83}, and recently explored on the lattice
\cite{Fiebig96,Lee97}.  

\section{INTERPOLATING FIELDS}

The excitation energies of the $1/2^-$ baryons can be extracted from
the standard two-point correlation function in the QCD vacuum
\begin{equation}
G({\bf p}, t) = \sum_{\bf x} e^{-i {\bf p} \cdot {\bf x}}
\langle \,  0 \mid \chi_{1/2}^-(x) \, \overline\chi_{1/2}^-(0) \mid 0 \, \rangle \,
\end{equation}
where $\chi_{1/2}^-$ is an odd-parity interpolating field used to
excite the hadron in question.  We consider odd-parity interpolating
fields based on the two established even-parity interpolating fields
for the nucleon ground state.  Left multiplication by $\gamma_5$
provides the following odd-parity interpolating fields
\begin{equation}
\chi_{N 1}^-(x) = \epsilon^{abc}
                 \left ( u^{Ta}(x) C \gamma_5 d^b(x) \right ) \gamma_5
                 u^c(x) \, ,
\end{equation}
\begin{equation}
\chi_{N 2}^-(x) = \epsilon^{abc}
                 \left ( u^{Ta}(x) C d^b(x) \right ) u^c(x) \, .
\end{equation}
The interpolating fields for other members of the $1/2^-$ octet are
obtained in a similar fashion \cite{dblOctet}.  A nonrelativistic
reduction of these interpolators reveals that $\chi_1^-$ pairs the
$u$-$d$ quark field operators in parentheses into a scalar diquark
while $\chi_2^-$ provides overlap with vector diquark pairs.  Products
of upper and lower components introduce derivatives of the quark field
operators.  Using SU(6) spin-flavor symmetry, the two nearby low-lying
$N^*1/2^-$ states of the physical spectrum are described by coupling
three quarks to a spin-1/2 or spin-3/2 spin-flavor wave function
coupled in turn to one unit of orbital angular momentum.  In a
relativistic field theory one does not expect $\chi_1^-$ and
$\chi_2^-$ to isolate individual states.  However, it is expected that
$\chi_1^-$ will predominantly excite the lower-lying $j=1/2$ state
associated with $\ell=1$ and $s=1/2$ in the quark model while
$\chi_2^-$ will predominantly excite the higher-lying state associated
with $s=3/2$.  This expectation is borne out in the following.

As for the even-parity $\Lambda$ interpolating fields \cite{dblOctet},
we consider two full octet interpolators $\chi_{\Lambda^O 1}^-$ and
$\chi_{\Lambda^O 2}^-$ and also consider two odd-parity interpolators
containing terms common to the flavor-octet and singlet $\Lambda$
interpolators $\chi_{\Lambda^C 1}^-$ and $\chi_{\Lambda^C 2}^-$
\cite{dblOctet}.  The latter two interpolating fields do not
strongly bias the flavor symmetry of the odd-parity $\Lambda$ state,
but rather allow the lowest lying odd-parity state to dominate the
correlator.

Despite having explicit negative-parity, these interpolating
fields couple to both positive and negative parity states.  
Hence, the parity of the intermediate state must be projected.

\section{PARITY PROJECTION}

The ability of the interpolating fields to couple to intermediate
states is described by a phenomenological parameter $\lambda$,
defined by
\begin{eqnarray}
\langle \,  0 \mid \chi_{1/2}^-(0) \mid N^+ \, \rangle &=&
\lambda_+ \, \gamma_5 \, u(p) \, , \\
\langle \,  0 \mid \chi_{1/2}^-(0) \mid N^- \, \rangle &=&
\lambda_- \, u(p), 
\end{eqnarray}
where $u(p)$ denotes a spinor.  
In the large Euclidean time limit
\[
G({\bf p}, t\to\infty) = \, e^{-E_\pm({\bf p}) \, t} \, \times
\]
\begin{equation}
\quad \sum_{s} 
\langle \,  0 \mid \chi_{1/2}^-(0) \mid N_s^\pm \, \rangle \langle \,
N_s^\pm \mid \overline\chi_{1/2}^-(0) \mid 0 \, \rangle \\
\end{equation}
\begin{equation}
= \lambda_+^2 { \gamma \cdot p - M_+ \over 2 M_+} e^{- E_+ \, t}
+   \lambda_-^2 { \gamma \cdot p + M_- \over 2 M_-} e^{- E_- \, t}
\label{eq:LargeT}
\end{equation}
where $p$ is on shell and $E_\pm$ is the on-shell energy $({\bf p}^2 +
M_\pm^2)^{1/2}$.  The large Euclidean time evolution of the two-point
function will ultimately be dominated by the lower-lying even-parity
state.  At zero momentum, even-parity states can be easily eliminated
by taking the trace of $G({\bf p}, t)$ with
\begin{equation}
\Gamma_4 \equiv  {1 \over 4} \left ( 1 + \gamma_4 \right )
= {1 \over 2} \left (
\begin{tabular}{ll}
I &0 \\
0 &0
\end{tabular}
\right ).
\end{equation}
Hence only the first two diagonal elements of the $4\times 4$ matrix
$G({\bf p}, t)$ are needed to isolate the mass of the odd-parity
state.  At finite momentum, the even-parity state can be
eliminated by taking the trace with
\begin{equation}
\Gamma_4({\bf p})\equiv {1 \over 4} \left ( 1 + {m_+ \over E_+} \gamma_4 
\right ).
\end{equation}
Note that this matrix remains diagonal such that only four elements of
$G({\bf p}, t)$ are needed to isolate the odd-parity state of
(\ref{eq:LargeT}) at finite momenta.  

Implicit in (\ref{eq:LargeT}) are two assumptions.  First the spectral
representation of the odd-parity resonance has been taken to be a
pole.  For the quark masses considered in this simulation, this is in
fact correct.  For example, the $\pi N$ decay channel of the low-lying
$N^*1/2^-$ is closed by energy conservation.  The issue of odd-parity
projection centers on the elimination of the ground-state pole.  As
such, the procedure outlined here continues to be valid as the
resonance decay channel opens upon decreasing the quark masses.

In addition, it is assumed there is only one positive parity state
lying below the lowest-lying $N^*1/2-$ state.  However, in the
physical nucleon spectrum there is both the ground state and the
$N^*1/2^+(1440)$ Roper resonance.  Since the mass of the Roper and the
odd-parity states of interest are quite close, the prefactors
of the exponentials in (\ref{eq:LargeT}) will govern the relative
contributions of these states over a wide range of Euclidean time.
For small momenta, (\ref{eq:LargeT}) indicates the Roper will
contaminate the projected $N^*1/2^-$ state at the level of a few
percent.  

\section{RESULTS}

\begin{table*}[t]
\setlength{\tabcolsep}{0.9pc}
\caption{Mass ratios of the $1/2^-$ baryons to the nucleon.  The
uncertainties are statistical. The three pion to rho mass ratios
correspond to quark masses of approximately 210, 180, 90 MeV.  The
assignment of the state labels is our conjecture.}
\label{ratio12}
\begin{tabular*}{\textwidth}{llccclc}
\hline
Interpolator &
Ratio  &
${\pi / \rho}=0.82$  &
${\pi / \rho}=0.78$  &
${\pi / \rho}=0.73$  & Chiral & Expt.\\
\hline
$\chi_{N 1}^-$ 
&${N / N(1535)}$ 
& 0.89(4)  & 0.85(5)  & 0.84(6) & 0.71(10)  & 0.61  \\
$\chi_{N 2}^-$ 
&${N / N(1650)}$
& 0.77(5)  & 0.71(6)  & 0.62(9) & 0.42(11)  &  0.57 \\
\hline
$\chi_{\Lambda^O 1}^-$ 
&${N / \Lambda(1670)}$
& 0.89(5)  & 0.85(5)  & 0.81(6) & 0.65(8)  &  0.56 \\
$\chi_{\Lambda^O 2}^-$ 
&${N / \Lambda(1800)}$
& 0.77(5)  & 0.71(6)  & 0.64(9) & 0.45(9)  &  0.52 \\
$\chi_{\Lambda^C 1}^-$ 
&${N / \Lambda(1405)}$
& 0.87(5)  & 0.83(6)  & 0.80(7) & 0.66(10)  & 0.67  \\
$\chi_{\Lambda^C 2}^-$ 
&${N / \Lambda(1670)}$
& 0.80(5)  & 0.76(6)  & 0.71(7) & 0.54(9)  &  0.56 \\
\hline
$\chi_{\Sigma 1}^-$ 
&${N / \Sigma(1620)}$
& 0.91(5)  & 0.85(5)  & 0.81(6) & 0.62(7)   &  0.58 \\
$\chi_{\Sigma 2}^-$ 
&${N / \Sigma(1750)}$
& 0.76(6)  & 0.71(6)  & 0.67(8) & 0.51(9)   &  0.54 \\
\hline
$\chi_{\Xi 1}^-$ 
&${N / \Xi(1620)}$
& 0.90(5)  & 0.85(5)  & 0.81(6) & 0.63(5)   &  0.58 \\
$\chi_{\Xi 2}^-$ 
&${N / \Xi(1690)}$
& 0.77(6)  & 0.71(6)  & 0.66(7) & 0.48(8)   &  0.56 \\
\hline
\end{tabular*}
\end{table*}

\begin{figure}[t]
\vspace{-5pt}
\begin{center}
\epsfxsize=6.0truecm
\leavevmode
\epsfbox{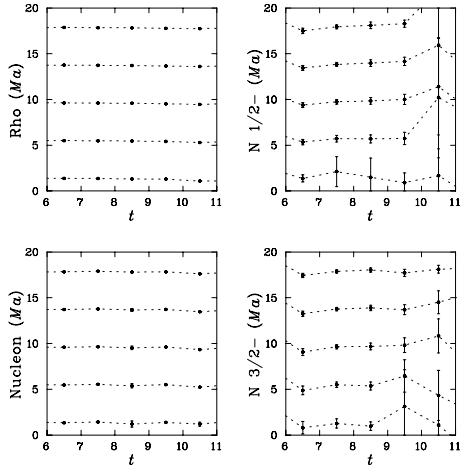}
\end{center}
\vspace{-30pt}
\caption{Effective hadron masses.  The five quark masses, heaviest
(top) to lightest (bottom), are offset for clarity.  The source is at
$t = 3$. }
\vspace{-20pt}
\label{effmass}
\end{figure}

We analyze 100 configurations on a $10^3 \times 16$ lattice at $\beta
= 7.0$ ($a \simeq 0.24$ fm).  Additional details of the simulation may
be found in Ref.\ \cite{Lee97}.  Figure \ref{effmass} displays
effective masses for the $N^*1/2^-$ and $N^*3/2^-$ states
\cite{3/2paper} compared to nucleon and rho masses.  A reasonable
plateau in the effective mass is seen for time slices 7 through 10 for
all five quark masses considered.

For the first qualitative look at the lattice data, the $1/2^-$ masses
are linearly extrapolated to the chiral limit.  
Mass ratios of the $1/2^-$ states to the nucleon mass are summarized
in Table~\ref{ratio12} at three quark masses and the chiral limit.
Clear separations of the odd-parity states from the ground state
nucleon are displayed.  Furthermore, it appears $\chi_1^-$ couples
predominantly to the ground state in the $1/2^-$ spectrum, while
$\chi_2^-$ mostly to the next state as anticipated.  This is further
supported by the fact that the off-diagonal correlation function
$(\chi_1^- \overline\chi_2^- + \chi_2^- \overline\chi_1^-)/2$ is
smaller by more than a factor of two compared to the diagonal
correlation functions $\chi_1^- \overline\chi_1^-$ or $\chi_2^-
\overline\chi_2^-$. 

%
We have also been investigating the odd-parity spin-3/2 nucleon
resonance \cite{3/2paper}.  The computational demand increases
dramatically relative to the proton.
 We are currently exploring 
 ways to reduce this factor.  
Preliminary analysis based on 25
configurations indicates a clear splitting from the ground state
nucleon, an encouraging result.

This exploratory study shows promise in extracting $N^*$ structure
from lattice QCD.  To improve the correlation function data for $N^*$
states, we plan to fine tune the source/sink smearing functions, use
anisotropic lattices with a fine lattice spacing in the
time-dimension, and substantially increase statistics.
 After the spectrum of the odd-parity states is understood, we plan to
 explore quantities of greater phenomenological interest, such as the
 $N \rightarrow N^*$ electromagnetic transition form factors.


\begin{thebibliography}{9}
\bibitem{Hamber83} H. Hamber and C.M. Wu,
Phys. Lett. {\bf B133} (1983) 351; {\bf B136} (1984) 255.

\bibitem{Fiebig96} H.R. Fiebig and R.M. Woloshyn,
Phys. Lett. {\bf B385} (1996) 273.

\bibitem{Lee97} F.X. Lee, D.B. Leinweber, hep-lat/9711044.

\bibitem{Chung82} Chung, Dosch, Kremer, Schall, Nucl. Phys. {\bf B197}
(1982) 55. 

\bibitem{dblOctet} D.~B. Leinweber, R.~M. Woloshyn, and T. Draper,
  Phys.\ Rev.\ D {\bf 43} (1991) 1659.

\bibitem{Leinweber95} D.B. Leinweber, Phys.\ Rev.\ D {\bf 51} (1995) 6383.


\bibitem{3/2paper} F.X. Lee, S. Choe and D.B. Leinweber.  Work in
progress. 

\end{thebibliography}
\end{document}